\documentclass[12pt]{iopart}
\usepackage{graphicx}
\usepackage{array}
\usepackage{url}
\usepackage{iopams}

\begin{document}

\title{Simulating streamer discharges in 3D with the parallel adaptive Afivo framework}

\author{Jannis Teunissen$^1$, Ute Ebert$^{2,3}$}

\address{$^1$Centre for mathematical Plasma Astrophysics, Department of
  Mathematics, KU Leuven, Celestijnenlaan 200B, B-3001 Leuven, Belgium}

\address{$^2$Centrum Wiskunde \& Informatica (CWI), P.O. Box 94079, 1090 GB
  Amsterdam, The Netherlands}

\address{$^3$Dept.\ Physics, Eindhoven Univ.\ Techn., The Netherlands}

\ead{jannis@teunissen.net}

\begin{abstract}
  We present an open-source plasma fluid code for 2D, cylindrical and 3D
  simulations of streamer discharges, based on the Afivo framework that features
  adaptive mesh refinement, geometric multigrid methods for Poisson's equation,
  and OpenMP parallelism. We describe the numerical implementation of a fluid
  model of the drift-diffusion-reaction type, combined with the local field
  approximation. Then we demonstrate its functionality with 3D simulations of
  long positive streamers in nitrogen in undervolted gaps, using three examples. The first example
  shows how a stochastic background density affects streamer propagation and
  branching. The second one focuses on the interaction of a streamer with preionized
  regions, and the third one investigates the interaction between two streamers. The
  simulations run on up to $10^8$ grid cells within less than a day. Without
  mesh refinement, they would require $4\cdot 10^{12}$ grid cells.
\end{abstract}

\maketitle


\section{Introduction}
\label{sec:introduction}

Streamer discharges~\cite{Vitello_1994,Yi_2002,Ebert_2010} are 
a generic stage of electric breakdown of nonconducting matter, 
dominated by strong space charge effects at the tips of growing discharge channels.
They occur as precursors of sparks, arcs and lightning leaders,
in nature as well as in high voltage and plasma technology.
Streamers are directly visible as so-called sprites in the mesosphere \cite{Sentman_1995},
and they are used in applications such as surface processing \cite{Cernak_2011},
sterilization and disinfection \cite{Akiyama_2000} or wound healing
\cite{Nastuta_2011}, often in the form of atmospheric pressure plasma jets
\cite{Sands_2008}.

\begin{figure}
  \centering
  \includegraphics[width=7cm]{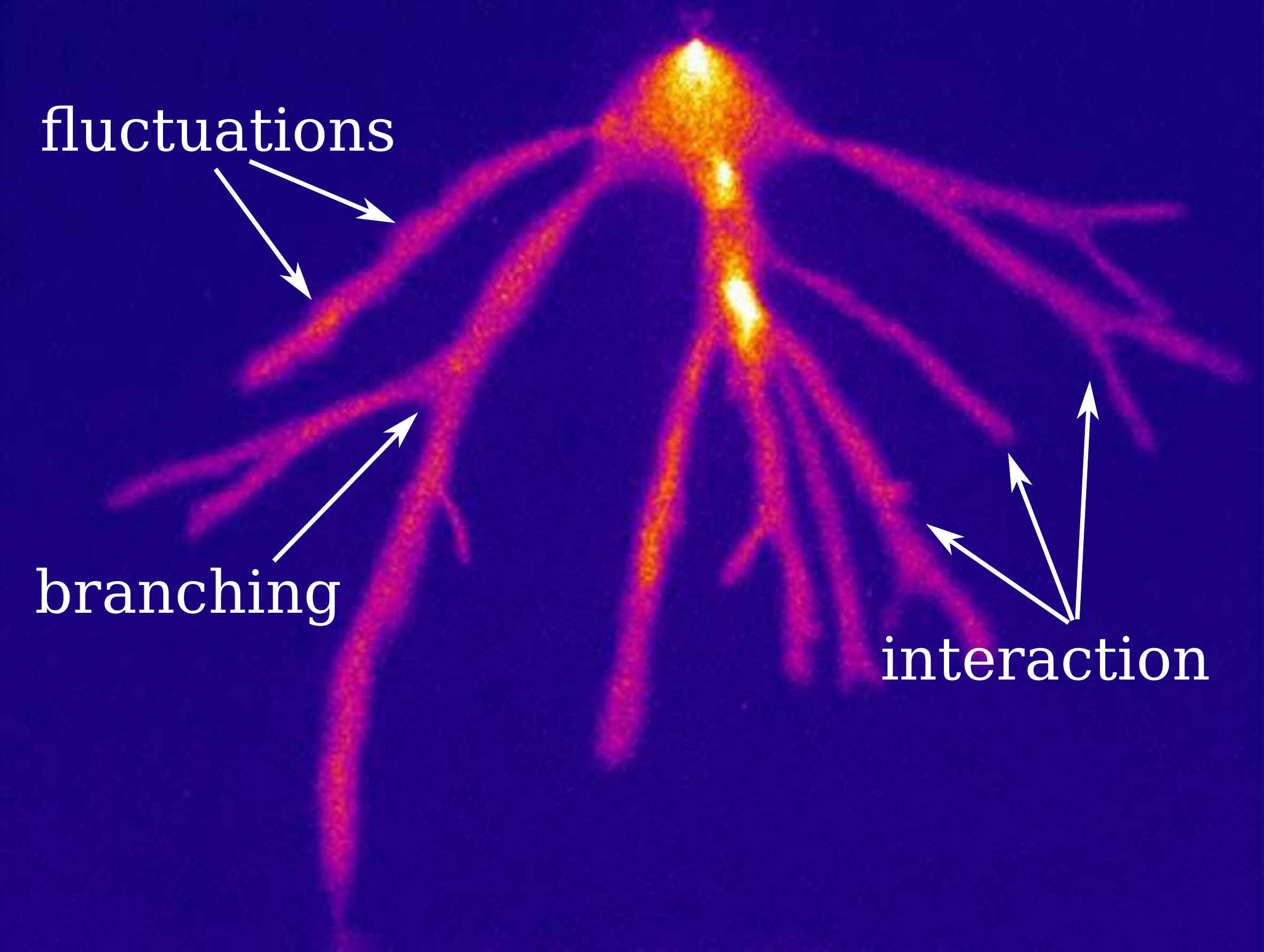}
  \caption{Experimental picture of positive streamers, showing why 3D simulations are
    often required: streamers branch and interact, and single streamers often
    show fluctuations which cannot be captured with axisymmetric models. Picture
    adapted from~\cite{Briels_2008_similarity} (air at 293~K and 0.4~bar,
	16~kV applied to a 4~cm gap).}
  \label{fig:motivation-3d}
\end{figure}

\begin{figure}
  \centering
  \includegraphics[width=0.9\textwidth]{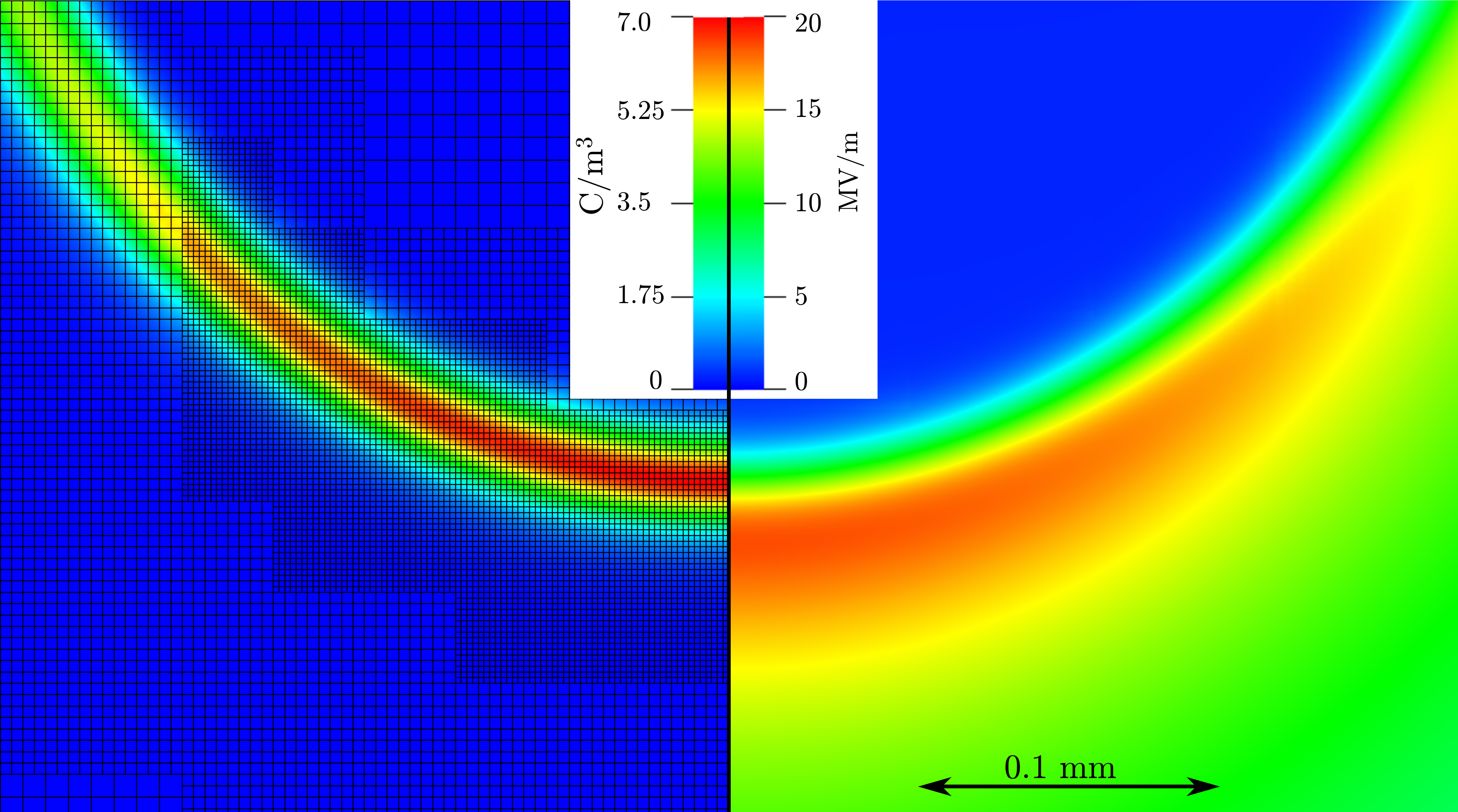}
  \caption{Cross section through a streamer head, showing the charge density
    (left) and electric field strength (right). The numerical mesh was generated
    according to the criteria described in section \ref{sec:ref-criterion},
    using equation (\ref{eq:ref-criterion}) with $c_0 = 1.0$ and $c_1 = 1.2$.}
  \label{fig:mesh-example}
\end{figure}

Streamers grow due to strong field enhancement at the tips of their partially 
ionized long channels. The high local fields support the 
local growth of ionization due to electron impact ionization. Simulating this
process has proven to be challenging for a number of reasons:
\begin{itemize}
  \item Problems such as streamer branching or the interaction between streamers
  require a three-dimensional description, as illustrated in figure
  \ref{fig:motivation-3d}.
  \item A fine grid spacing is required to accurately resolve the thin charge
  layers around streamer heads that create the local field enhancement, see
  figure \ref{fig:mesh-example}. Due to the strongly non-linear growth of
  streamers, it is usually not possible to obtain an approximate solution on a
  coarse grid.
  \item Time-dependent simulations are required. Due to the high electric field
  at streamer tips, where the mesh spacing is small, small time steps have to be
  used.
  \item At each time step, Poisson's equation has to be solved to obtain the
  electrostatic potential and field. The non-local nature of this equation
  complicates the parallelization of streamer models.
\end{itemize}
The physics of streamer discharges is mostly governed by electrons, because
ions gain energy more slowly and lose it more easily in collisions. Both plasma
fluid models and kinetic/particle-in-cell models have been used to simulate
streamers. In fluid models particle densities (and sometimes also momentum or
energy densities) evolve in time, using pre-calculated transport coefficients as
input data. In kinetic simulations, the electron distribution function
$f(\vec{x}, \vec{v}, t)$ evolves in time, using cross sections as input data.
Kinetic simulations typically require a large number of particles and smaller
time steps than fluid simulations, so that their computational cost is
considerably higher.

The first demonstration of a 3D fluid simulation was given
in~\cite{Kulikovsky_1998b}. In~\cite{Pancheshnyi_2008}, parallel fluid
simulations with adaptive mesh refinement (AMR) were performed using Paramesh,
but the main bottleneck was the Poisson solver. Later work includes a 2.5D fluid
model with AMR, parallelized over axial modes~\cite{Luque_2008_b}.
Kinetic~\cite{Chanrion_2008,Rose_2011} and hybrid
kinetic/fluid~\cite{Li_hybrid_ii_2012} 3D models without AMR have also been
employed, and more recently kinetic models with AMR have been
used~\cite{Teunissen_2016,Kolobov_2016}. Other work includes 3D simulations with
a finite element code~\cite{Papageorgiou_2011} and a proof-of-concept of 3D
simulations in transformer oil with OpenFoam~\cite{Lavesson_2014}. A notable
development was the 3D fluid model with AMR presented in~\cite{Kolobov_2012},
which is commercially available. The adaptation of parallel multigrid methods
from the Gerris Flow Solver~\cite{Popinet_2003} made it possible to perform
relatively large scale 3D simulations.

Here we present Afivo-streamer, an open-source fluid model for the simulation of
streamer discharges. Both 2D, axisymmetric and 3D simulations are supported, but
the focus here is on 3D, which is computationally most challenging.
Afivo-streamer is based on the Afivo\footnote{Afivo stands for ``Adaptive Finite
  Volume Octree''} framework \cite{Teunissen_afivo_arxiv}, which provides
quadtree/octree adaptive mesh refinement, a geometric multigrid solver,
shared-memory parallelism, and routines for writing output. A first successful
application of Afivo-streamer can be found in~\cite{Nijdam_Teunissen_2016}. The
main contribution of Afivo-streamer is that it provides efficient and
open-source computational infrastructure for 2D, 3D and axisymmetric streamer
simulations. The paper is organized in two parts. In the first part (section
\ref{sec:model}), the numerical implementation of Afivo-streamer is described.
In the second part (section \ref{sec:3d-simulations}), we demonstrate the code's
functionality with three 3D examples.

\section{Model description}
\label{sec:model}

The implementation of the different components of Afivo-streamer is described
below. The source code is available online through \cite{cwimd} under an open
source (GPLv3) license.

\subsection{Afivo AMR framework}
\label{sec:Afivo}

Adaptive mesh refinement is essential for 3D streamer simulations. Without AMR,
the fine grid spacing that is required near the streamer head severely restricts
the size of the computational domain. Here the open-source Afivo framework
\cite{Teunissen_afivo_arxiv} is used to provide AMR and parallelization for
streamer simulations. The functionality of Afivo is summarized below; for more
details we refer to~\cite{Teunissen_afivo_arxiv}.

\subsubsection{Adaptive quadtree/octree grids}

\begin{figure}
  \centering
  \includegraphics[width=0.9\textwidth]{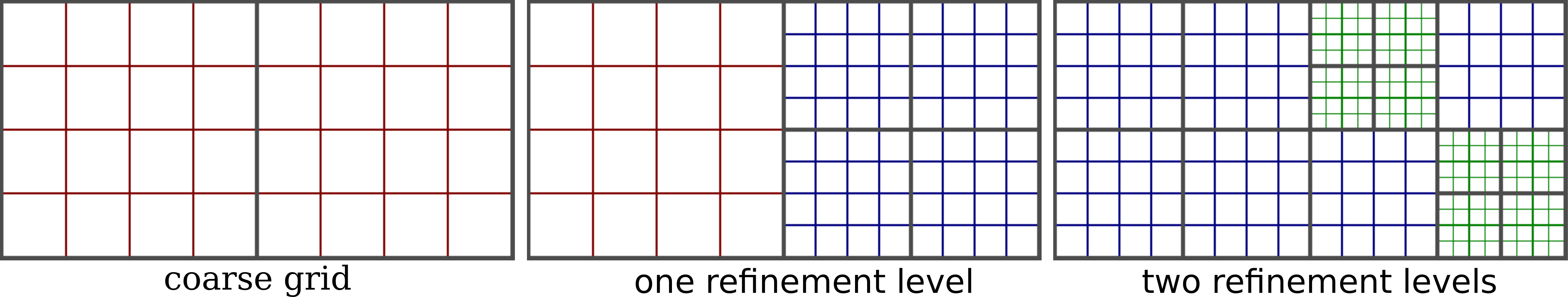}
  \caption{Left: example of a quadtree grid consisting of two blocks of
    $4 \times 4$ cells. The middle and right figure show how the mesh can be
    refined by recursively adding new blocks, each having half the grid spacing
    of their parent. 
	}
  \label{fig:quadtree-comp-domain}
\end{figure}

Afivo supports quadtree (2D) and octree (3D) grids. A quadtree/octree grid
consists of blocks of $N^D$ cells, where $N$ is an even number (here we use
$N=8$) and $D$ is the problem dimension. One or more of these blocks defines the
coarse grid. A coarse grid block can be refined by covering it with $2^D$ child
blocks, which each have half the grid spacing. This process can be repeated
recursively, leading to an adaptively refined mesh that still has a quite
regular structure, as illustrated in figure \ref{fig:quadtree-comp-domain}.

Afivo provides routines for adapting the mesh, but does not come with built-in
refinement criteria. The criteria used here are discussed in section
\ref{sec:ref-criterion}. Afivo does ensure \emph{proper nesting}, which means
that neighboring boxes differ by at most one refinement level. Methods for
interpolating from coarse to fine grids, and vice versa, are included.

\subsubsection{Geometric multigrid solver}
\label{sec:geom-mult-solv}

One of the key computational challenges in streamer simulations is quickly
solving Poisson's equation
\begin{equation}
  \label{eq:poisson}
  \nabla \cdot (\varepsilon\nabla\phi) = -\rho,
\end{equation}
to obtain the electrostatic potential $\phi$ from the charge density $\rho$,
where $\varepsilon$ is the dielectric permittivity. The electrostatic field can
then be determined as $\vec{E} = -\nabla \phi$. Poisson's equation has to be
solved at every time step and with high spatial resolution within the ionization fronts, 
and its non-local nature prevents a straightforward
parallel solution. Therefore, the Poisson solver is often the most
time-consuming part of streamer simulations. Afivo implements geometric
multigrid routines~\cite{Brandt_2011,Trottenberg_2000_multigrid}, which are among the fastest methods for solving
elliptic equations such as (\ref{eq:poisson}).

Multigrid methods are iterative solvers which cycle over a hierarchy of grids.
Short-wavelength errors are efficiently reduced on fine grids, and
long-wavelength errors on coarse grids, by using an appropriate smoothing
procedure. There are many varieties of multigrid, which differ in for example
their multigrid cycle, smoothing procedure, grid hierarchy or interpolation
method. For a detailed description of multigrid methods, which we cannot give
here, we refer to
e.g.~\cite{Brandt_2011,Trottenberg_2000_multigrid,Briggs_2000}.

Afivo supports a V-cycle and an FMG (full multigrid) cycle. An FMG cycle is more
expensive than a V-cycle, but it typically gives a solution within the
discretization error in one or two iterations. Both cycles are implemented using
the full approximation scheme, which means that the computed solution is
available at all grid levels (in some multigrid methods, only the correction to
the solution is computed on coarse grids). Afivo includes Gauss-Seidel red-black
smoothers that can be used for constant $\varepsilon$ problems, and to some
extent also for problems where $\varepsilon$ varies,
see~\cite{Teunissen_afivo_arxiv}. It is currently not possible to include
internal boundary conditions, for example to define a curved electrode, although
work in that direction is ongoing.

For the Afivo-streamer model, an FMG cycle is used to compute the initial
electric potential. For each subsequent update of the potential a number of
V-cycles is used (here two), which use the previous solution as an initial
guess. This exploits the fact that there are only small changes in the potential
between time steps.

In streamer simulations with AMR, the fine grid ideally covers a relatively
small region. As discharges propagate, the mesh has to follow their features,
meaning it changes frequently in time. A key advantage of \emph{geometric}
multigrid methods is that they require almost no extra computation when the mesh
changes, in contrast to matrix-based (direct) methods.

\subsubsection{Parallelization } 
\label{sec:parallelization} 

Afivo incorporates shared-memory parallelization using OpenMP, which means that
it can use one up to e.g. 32 cores, depending on the available hardware. Since the quadtree/octree grid is naturally
divided into blocks, the parallelization is performed over these blocks. Each
block contains a layer of ghost cells, so that they can be operated on
independently. The scaling of codes based on Afivo is typically limited by the
memory bandwidth of the computer. For the multigrid methods, a further challenge
is that coarse grid levels contain few data points, hampering their parallel
efficiency, see e.g. \cite{Trottenberg_2000_multigrid}.

\subsubsection{Writing output}
\label{sec:writing-output}

When doing 3D simulations with AMR, writing and visualizing output can be
challenging. Afivo comes with support for writing VTK unstructured files and
Silo files, which can be visualized with e.g. Visit~\cite{HPV:VisIt}. For 3D
simulations, the Silo format is more efficient, as it groups grid blocks into
larger rectangular regions. The Silo files also include ghost cell information,
which helps to ensure smooth visualizations near refinement boundaries. For a 3D
streamer simulation, output can get pretty large: using for example 5 variables
and $2 \times 10^7$ grid cells, a single file is about a gigabyte.

\subsection{Fluid model equations}
\label{sec:fluid-model-equat}

The fluid model used here is of the drift-diffusion-reaction type with the local
field approximation~\cite{Luque_2012}. It keeps track of the electron density
$n_e$ and the positive ion density $n_i$:
\begin{eqnarray}
  \label{eq:a5-fluid-model}
  \partial_t n_e &= \nabla \cdot (\mu_e n_e \vec{E} + D_e \nabla n_e) +
                   \bar{\alpha} \mu_e E n_e,\\
  \partial_t n_i &= \bar{\alpha} \mu_e E n_e.
\end{eqnarray}
Here $\bar{\alpha}$ is the effective ionization coefficient, $\mu_e$ the
electron mobility, $D_e$ the electron diffusion coefficient and $\vec{E}$ the
electric field. With the \emph{local field approximation} $\mu_e$, $D_e$ and
$\bar{\alpha}$ are functions of the local electric field strength. These
coefficients can be computed with a Boltzmann solver
\cite{Hagelaar_2005,Dujko_2011} or particle swarms~\cite{Li_hybrid_i_2010}, or
they can be measured experimentally. The fluid equations are coupled to the
electrostatic field, which is computed as
\begin{eqnarray}
  \vec{E} &= -\nabla \phi,\\
  \nabla^2 \phi &= -e (n_i - n_e)/\varepsilon_0
\end{eqnarray}
where $\phi$ is the electric potential, $\varepsilon_0$ the permittivity of
vacuum and $e$ the elementary charge. The electric potential is computed with
the multigrid routines from Afivo, described in section \ref{sec:Afivo}.

Different types of plasma fluid models can be implemented in Afivo-streamer.
More advanced models could for example include an equation for the momentum
and/or energy density, and let the transport coefficients depend on the mean
electron energy, see e.g.~\cite{Markosyan_2015,Becker_2017}. The mean energy is
then given by $Q/n_e$, where $Q$ is the energy density. Such models capture more
of the physics, as demonstrated in e.g. \cite{Eichwald_2005}. However, the ratio
is $Q/n_e$ hard to define when $n_e \to 0$, making such models less robust than
the one used here. Furthermore, a hyperbolic system with multiple coupled
equations is generally harder to solve than a scalar one.

For electric discharges in air, photoionization is often an important process
\cite{Pancheshnyi_2014}. Excited nitrogen molecules can emit UV photons which
are able to ionize oxygen molecules. Such a non-local source of free electrons
is particularly important for positive streamers, which require free electrons
ahead of them to grow. Afivo-streamer contains a Monte Carlo procedure for
photoionization, which can take into account stochastic fluctuations due the
finite number of photons. The procedure is described in chapter 11 of
\cite{Teunissen_thesis_2015}, and in a forthcoming paper we will investigate the
effect of stochastic photoionization on streamer branching. In the present paper
we focus on discharges in pure nitrogen without photoionization, using a
background density of electrons and positive ions.

\subsection{Spatial discretization}
\label{sec:spatial-discr}

The spatial discretizations used in Afivo-streamer are second order accurate. We
use a finite volume approach, in which the following quantities are defined at
cell centers: the electron/ion density, the electric potential, and the electric
field strength. The electron fluxes and the electric field components are
defined at cell faces.

Afivo's multigrid routines compute the electric potential from the charge
density, as discussed in section \ref{sec:geom-mult-solv}. From the
cell-centered electric potential $\phi$, the electric field at cell faces is
computed by central differencing, so that the $x$-component is computed as
$$E_{x}^{i+1/2,j,k} = (\phi^{i,j,k} - \phi^{i+1,j,k}) / \Delta x.$$ The electric
field strength at cell centers is then computed as
$E^{i,j,k} = \sqrt{E_x^2 + E_y^2 + E_z^2}$ where
$E_x = (E_{x}^{i-1/2,j,k} + E_{x}^{i+1/2,j,k})/2$ is the average $x$-component
at the cell center, $E_y = (E_{y}^{i,j-1/2,k} + E_{y}^{i,j+1/2,k})/2$, and
similar for $E_z$.

We follow the approach from~\cite{Montijn_2006} for the discretization of the
fluid equations. The advective part of the flux is computed using the Koren
limiter~\cite{koren_limiter}. The electron velocity at a cell face is then computed
as $$v_{x}^{i+1/2, j, k} = -\mu(E^{*}) E_{x}^{i+1/2, j, k},$$ where
$E^{*} = (E^{i,j,k} + E^{i+1,j,k})/2$ is the electric field strength $|\vec{E}|$
at the cell face. For brevity, we now omit the extra indices $j,k$. If
$v_{x}^{i+1/2} < 0$, the advective flux between cell $i$ and $i+1$ is given by
\begin{equation}
  \label{eq:koren-neg}
  f_{x}^{i+1/2} = v_{x}^{i+1/2} \left(n_e^{i+1} -
  \psi\left(\frac{n_e^{i+2} - n_e^{i+1}}{n_e^{i+1} - n_e^{i}}\right)
  (n_e^{i+1} - n_e^{i})\right),
\end{equation}
and if $v_{x}^{i+1/2} \geq 0$, it is given by
\begin{equation}
  \label{eq:koren-pos}
  f_{x}^{i+1/2} = v_{x}^{i+1/2} \left(n_e^{i} +
  \psi\left(\frac{n_e^{i} - n_e^{i-1}}{n_e^{i+1} - n_e^{i}}\right)
  (n_e^{i+1} - n_e^{i})\right),
\end{equation}
where $\psi(x)$ is the Koren limiter, given by
\begin{equation*}
  \psi(x) = \max\left(0, \min(1, (2+x)/6, x)\right).
\end{equation*}
The $y$ and $z$ components are computed similarly. Note that the above
equations, if directly implemented, could cause division by zero. Our numerical
implementation avoids this; it is described in Appendix B of
\cite{Teunissen_thesis_2015}.

The diffusive flux between cells $i$ and $i+1$ is computed using
central differences, and is given by
\begin{equation*}
  f_x^{i+1/2} = D_e(E^{*}) (n_e^{i+1} - n_e^{i})/\Delta x,
\end{equation*}
with $E^{*}$ defined as above. 

To efficiently look up transport coefficients we
convert them to a \emph{lookup table}. This table stores the coefficients at
regularly spaced electric field strengths, linearly interpolating the input
data, e.g., from BOLSIG+. To look up values for a given field strength, 
the corresponding index in the table is
computed, after which linear interpolation is employed. By default, the table is
constructed up to $E_\mathrm{max} = 35 \, \textrm{MV/m}$ using 1000 entries.

Near refinement boundaries, we use linear interpolation to obtain two fine-grid
ghost values, which are required for equations
(\ref{eq:koren-neg}-\ref{eq:koren-pos}). These ghost cells lie inside a
coarse-grid neighbor cell, and we limit them to twice the coarse value to
preserve positivity\footnote{In the future, we intend to use a more general
  limiting procedure near refinement boundaries.}. At refinement boundaries, the
coarse fluxes are set to the sum of the fine fluxes to ensure mass conservation.

\subsection{Temporal discretization}
\label{sec:temporal-discr}

Time stepping is performed as in~\cite{Montijn_2006}, using the second order
accurate explicit trapezoidal rule. This method is strong stability preserving
(SSP) and has favorable properties when combined with the Koren limiter
\cite{Hundsdorfer_2003}. Our implementation advances over $\Delta t$ as follows:
\begin{enumerate}
  \item Store the original electron and ion densities.
  \item Compute fluxes and source terms, then perform a forward Euler step
  over $\Delta t$ and compute a new electric field.
  \item \label{enum:step2} Compute fluxes and source terms, then perform another
  forward Euler step over $\Delta t$.
  \item Average the new electron and ion densities (advanced over $2 \Delta t$)
  with the stored initial ones. Then compute a new electric field from the
  resulting charge density.
\end{enumerate}

All the grids are advanced using the same global time step. We limit $\Delta t$
according to several criteria. The first is a CFL condition
\begin{equation*}
  \Delta t \sum{|v_i|/\Delta x} < 0.5
\end{equation*}
where $v_i$ are the velocity components and $\Delta x$ the grid spacing. This
condition is more strict than necessary for stability, but we found that a CFL
number of $0.5$ gives a good balance between accuracy and computational cost. To
ensure stability for the combined advective and diffusive fluxes, we require
\begin{equation*}
  \Delta t \sum{|v_i|/\Delta x} + \Delta t \, (2 D D_e) / \Delta x^2 < 1.0,
\end{equation*}
where $D$ is the problem dimension, and $D_e$ the electron diffusion constant.
Finally, the time step is also limited by the dielectric relaxation time
\begin{equation*}
  \Delta t < \varepsilon_0 / (e \mu_e n_e).\\
\end{equation*}
These requirements for $\Delta t$ are evaluated at stage (\ref{enum:step2}) of
our time stepping scheme, where the required quantities are already available.
The next time step is then obtained by multiplying with a safety factor (default
$0.9$).

\subsection{Refinement criterion}
\label{sec:ref-criterion}

\begin{figure}
  \centering
  \includegraphics[width=8cm]{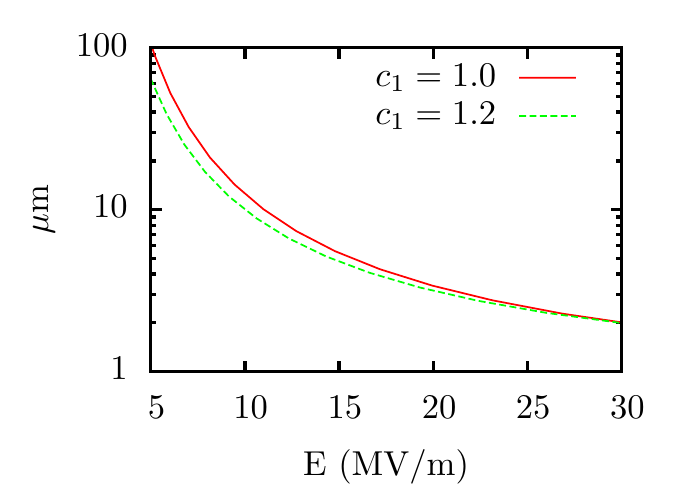}
  \caption{The refinement criterion of equation (\ref{eq:ref-criterion}) for
    $c_1 = 1.0$ and $c_1 = 1.2$, with $c_0 = 1.0$ for both cases. With a larger
    value of $c_1$, there is more refinement at low to intermediate electric
    fields. Data for nitrogen at 1 bar was used for $\alpha(E)$, as described in
    section \ref{sec:simulation-conditions}.}
  \label{fig:alpha-refinement}
\end{figure}

The growth of positive streamers is dominated by electron impact ionization.
Therefore, our refinement criterion is based on $1/\alpha(E)$, which is the
average distance between ionization events for an electron. Ignoring advection,
it is an estimate for the distance over which the electron density increases by
a factor of $e \approx 2.72$. For the simulations presented here, the following
criterion was used
\begin{equation}
  \label{eq:ref-criterion}
  \Delta x < c_0 c_1 / \alpha(c_1 E),
\end{equation}
where we used $c_0 = 1$ and $c_1 = 1.2$. The constant $c_1$ was introduced to
balance the refinement ahead and on the sides of the streamer. Without this
constant (or when it is one), we sometimes observed oscillations in a streamer's
radius. Setting $c_1 > 1$ increases the refinement for intermediate electric
fields, as illustrated in figure \ref{fig:alpha-refinement}. This helps to have
more refinement on the sides of streamers, without significantly increasing the
refinement at their tips.

The criterion of equation (\ref{eq:ref-criterion}) is evaluated for each grid
cell. If at least one cell of a grid block requires refinement, the whole block
is refined. Afivo implements a refinement buffer, so that blocks are also
refined when nearby cells in neighboring blocks require refinement. For the
simulations presented here, we used a buffer distance of three cells.
Furthermore, the code places refinement around the initial conditions, to ensure
they are accurately captured. Grid blocks can be derefined when for all cells
equation (\ref{eq:ref-criterion}) holds with $c_0 = 1/8$ and
$\Delta x < \Delta x_\mathrm{deref}$. Here, we used
$\Delta x_\mathrm{deref} = 30 \, \mu\textrm{m}$, which controls the mesh
resolution of the discharge in regions where it no longer grows. An example of
the resulting mesh around a streamer head is shown in figure
\ref{fig:mesh-example}. This streamer was generated in nitrogen at
$1 \, \mathrm{bar}$, using the same transport coefficients as for the examples
presented in section \ref{sec:3d-simulations}.

The above criterion is an empirical criterion for positive streamers, which
often works quite well, but not always. For example, when simulating negative
streamers propagating into a zero-density region ($n_e = 0$), the criterion will
trigger refinement where there are no electrons and no space charge. We have
experimented with a different criterion, based on the space charge density
$\rho$: $\Delta x < \sqrt{c_3 \varepsilon_0 / |\rho|}$, with $c_3$ for example
$25 \, \textrm{V}$. Such a criterion captures the space charge layers quite
well, but not the strong density gradients ahead of those charge layers, which
play an important role in streamer propagation. In the future, we hope to find a
more generic criterion, based on the discretization error in the model itself.

We would like to point out that the coarse mesh can make a significant
difference in the computational cost of simulations. For example, if the finest
mesh spacing required in a simulation is $2 \, \mu\textrm{m}$, and the
computational domain measures $(10 \, \textrm{mm})^3$, then the actual finest
mesh will have a spacing of about
$1.22 \, \mu\textrm{m} = 10/2^{13} \, \textrm{mm}$. By using a larger or smaller
computational domain, the fine-grid spacing can be made to agree better with its
desired value. This would allow for larger time steps, often using a smaller
total number of grid cells.

\section{3D simulations}
\label{sec:3d-simulations}

We now demonstrate the functionality of Afivo-streamer with three examples, all
in 3D. The simulations were performed on a single node containing two Xeon
E5-2680v4 processors ($2 \times 14$ cores, at $2.4 \, \textrm{GHz}$). The
simulations ran for up to 24 hours, using up to $10^8$ grid cells. Individual
output files with the 3D data were up to $5 \, \textrm{gigabyte}$ in size.

\subsection{Simulation conditions}
\label{sec:simulation-conditions}

The simulations presented here were performed in nitrogen at one bar. Electron
transport coefficients (e.g., $\alpha$, $\mu_e$) were computed with
Bolsig+~\cite{Hagelaar_2005} from Phelps' cross sections~\cite{Phelps_1985}. A
computational domain of $(40 \, \textrm{mm})^3$ was used, constructed from
octree blocks of $8^3$ cells. The maximum grid spacing was set to
$625 \, \mu\textrm{m}$; the minimum grid spacing in the simulations was about
$2.4 \, \mu\textrm{m}$. A background electric field of
$E_0 = 2.0 \, \textrm{MV/m}$ was applied in the $-\hat{z}$ direction, which is below the `breakdown'
threshold for nitrogen. For a discussion of the difference between discharges
in overvolted and undervolted conditions we refer to~\cite{Sun_2014}. The
background field is imposed by grounding the bottom
boundary of the domain and applying $80 \, \textrm{kV}$ at the top. On the other
sides of the domain, Neumann zero boundary conditions were used for the
potential. Neumann zero boundary conditions were also used for the electron
density on all sides, but this had little effect on the results because the
simulated streamers did not connect to boundaries.

The propagation of positive streamers requires free electrons ahead of them. In
air, such electrons are often provided by photoionization. Since we here perform
simulations in nitrogen, where photoionization is absent, a background density
of $10^{14} \, \textrm{m}^{-3}$ electrons and positive ions is included instead.
Such a density could for example be present due to previous discharges in a
repetitively pulsed system~\cite{Nijdam_2014_double_pulse}.

To start a discharge, the background field has to be locally enhanced. We do
this by placing an ionized seed of about $1.8 \, \textrm{mm}$ long with a radius
of about $0.15 \, \textrm{mm}$. The electron and positive ion density are
$10^{20} \, \textrm{m}^{-3}$ at the center, which decays at distances above
$d = 0.1 \, \textrm{mm}$ with a so-called \emph{smoothstep} profile:
$1 - 3x^2 + 2 x^3$, where $x = (d-0.1 \, \textrm{mm})/0.1 \, \textrm{mm}$. When
the electrons from a seed drift upwards, the electric field at the bottom of the
seed is enhanced so that a positive streamer can form.

\subsection{Stochastic background density}
\label{sec:stochastic-background}

\begin{figure}
  \centering
  \includegraphics[width=8cm]{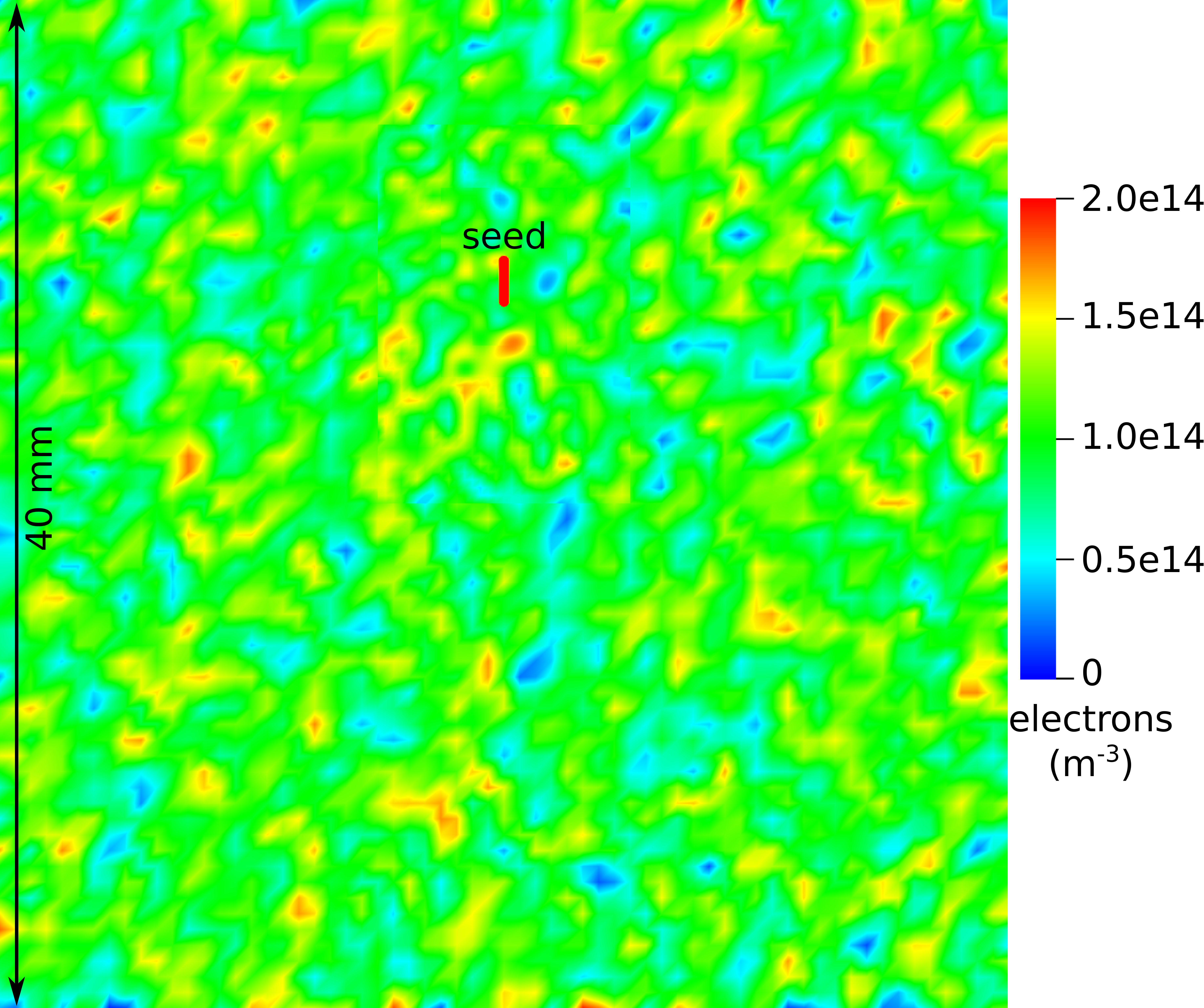}
  \caption{Cross section through the computational domain for case 3, showing a
    stochastic background density $2U \times 10^{14} \, \textrm{m}^{-3}$ with a
    correlation length of $625 \, \mu\textrm{m}$, where $U$ is a uniform random
    number between zero and one. The location of the ionized seed from which the
    discharge starts is also visible, its density ($10^{20} \, \textrm{m}^{-3}$)
    exceeds the color scheme.}
  \label{fig:stochastic-density}
\end{figure}

In this example we investigate how a stochastic distribution of background
ionization affects streamer propagation. A single ionized seed is placed as
shown in figure \ref{fig:stochastic-density}. We then let a discharge evolve
using three different background ionization distributions, for which the
electron and positive ion density per cell are given by:
\begin{itemize}
  \item Case 1: A constant value of $10^{14} \, \textrm{m}^{-3}$
  \item Case 2: A stochastic value $(0.5 + U) \times 10^{14} \, \textrm{m}^{-3}$, where
  $U$ is a uniformly distributed random number between zero and one.
  \item Case 3: A stochastic density $2U \times 10^{14} \, \textrm{m}^{-3}$,
  using the same random numbers as for case 2.
\end{itemize}
The background is created at the grid level with spacing $625 \, \mu\textrm{m}$
and then linearly interpolated to finer grids, so that the noise has a
correlation length of $625 \, \mu\textrm{m}$. Note that all three cases have the
same average density of $10^{14} \, \textrm{m}^{-3}$. An example of the third
case is shown in figure \ref{fig:stochastic-density}. We remark that the above
distributions do not contain physically realistic fluctuations, in which case
the number of electrons per cell would be Poisson-distributed.

\begin{figure}
  \centering
  \includegraphics[width=10cm]{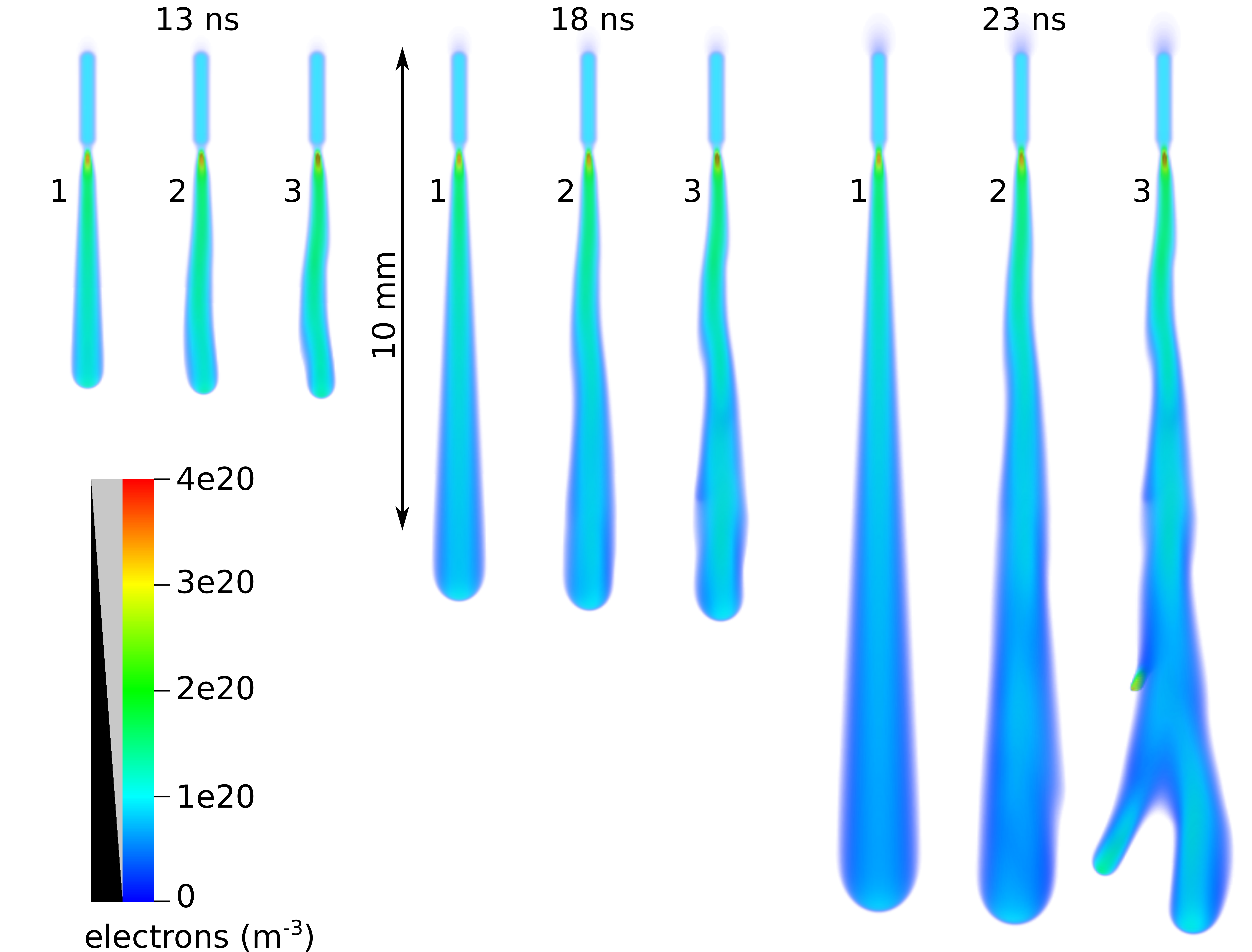}
  \caption{Evolution of a positive streamer in three background densities (1:
    uniform, 2: half-stochastic, 3: fully stochastic, see text). The average
    background density is $n_e = n_i = 10^{14} \, \textrm{m}^{-3}$ for each
    case. Shown is a 3D volume rendering of the electron density; the opacity is
    indicated in the legend.}
  \label{fig:stochastic-result}
\end{figure}

Figure \ref{fig:stochastic-result} shows how a positive streamer propagates for
the different cases. Remarkably, the streamer velocity is nearly identical. This
is consistent with previous studies
\cite{Teunissen_2016,Nijdam_2011,Wormeester_2010}, in which it was found that
the streamer velocity only weakly depends on the background ionization level.
The background density has a stronger effect on the morphology of the streamer.
After $23 \, \textrm{ns}$, case 3 shows streamer branching, while case 1 and 2
do not. The evolution of cases 2 and 3 seems closer to the experimentally
observed streamers of figure \ref{fig:motivation-3d}. Our results agree with a
previous study~\cite{Luque_2011_density_fluctuations}, in which it was found
that positive streamer branching is accelerated by stochastic electron density
fluctuations.

\subsection{Interaction with preionization}
\label{sec:ionized-regions}

\begin{figure}
  \centering
  \includegraphics[width=\textwidth]{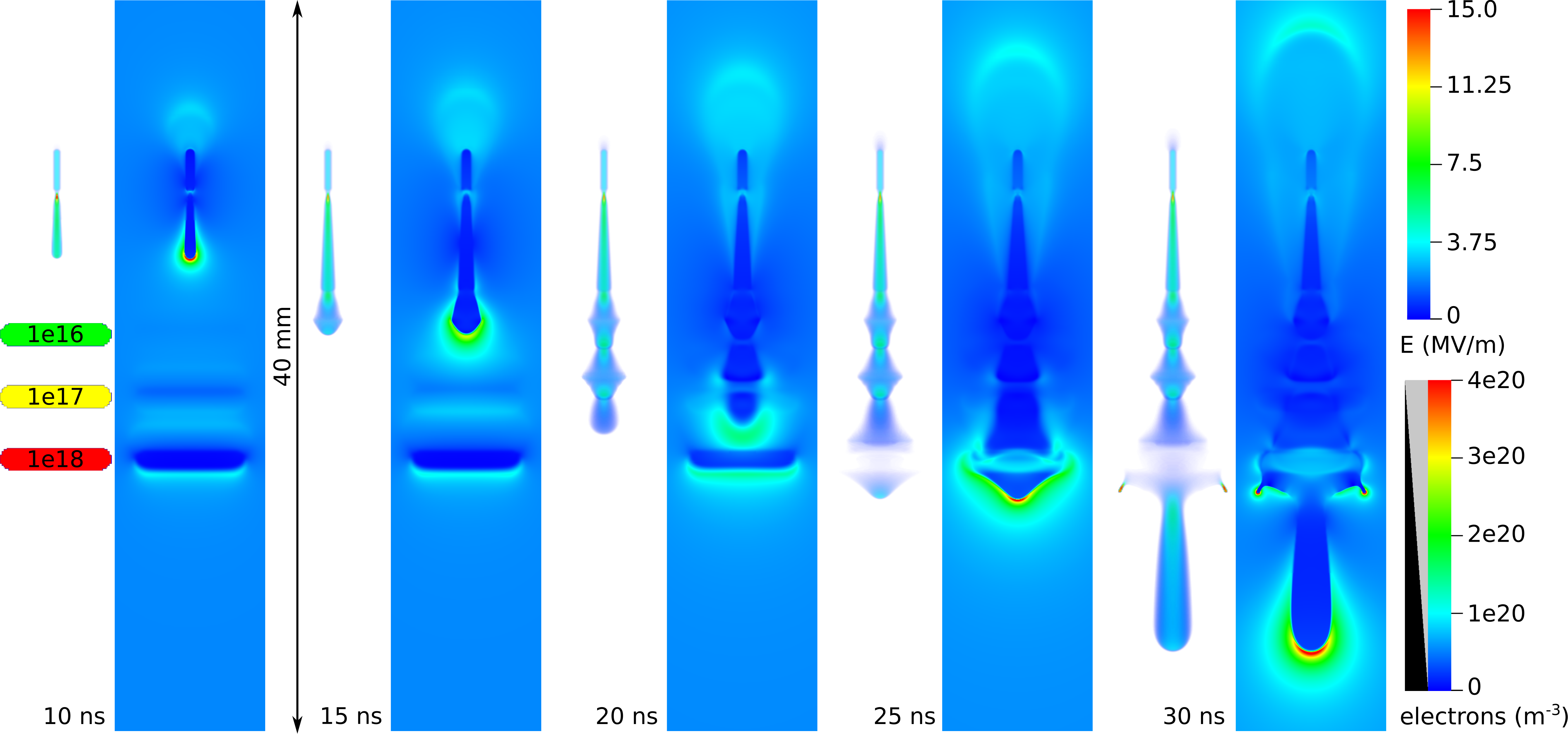}
  \caption{Evolution of a positive streamer as it propagates through preionized
    regions with $10^{16}$, $10^{17}$ and $10^{18} \, \textrm{m}^{-3}$ electrons
    and positive ions. At each indicated time a 3D volume rendering of the
    electron density together with a cross section of the electric field is
    shown.}
  \label{fig:stripes-result}
\end{figure}

This example is related to two previous studies
\cite{Nijdam_Teunissen_2016,Nijdam_2014}, in which the guiding of positive
streamers by preionization from a laser was investigated. Here, we simulate a
positive streamer passing through three preionized cylinders. The cylinders are
aligned perpendicular to the direction of propagation, as indicated on the left
of figure \ref{fig:stripes-result}. They contain a density of $10^{16}$,
$10^{17}$ and $10^{18} \, \textrm{m}^{-3}$ electrons and positive ions. A
background density of $10^{14} \, \textrm{m}^{-3}$ was present in the whole
domain.

Figure \ref{fig:stripes-result} shows how the electron density and electric
field evolve in time. Since the fluid model employed here is deterministic, the
left-right symmetry in the initial conditions is preserved. Upon reaching the
first preionized region, the streamer's maximum electric field is reduced, and
it becomes slightly wider. The second patch has a similar, but somewhat stronger
effect. Inside the third patch, the streamer temporarily disappears, at least
when looking at the electron density. Due to the high preionization density
($10^{18} \, \textrm{m}^{-3}$) in this region, the streamer loses most of its
electric field enhancement. A similar phenomenon was observed for
sprite discharges, to explain the formation of so-called `beads'
\cite{Luque_2011b}. At around $25 \, \textrm{ns}$ the streamer continues, and two
branches form at the boundary of the preionized cylinder. As the positive
streamer grows downwards, electrons drift out towards the top. These electrons
could eventually form a negative streamer, as can be seen in the electric field
profiles at later times.

\subsection{Interacting streamers}
\label{sec:two-parallel}

In this example the interaction between two streamers is investigated;
previous numerical and experimental investigations can be found in
\cite{Luque_2008_b,Nijdam_2009}.
The two interacting streamers are created by placing two field-enhancing seeds in the domain,
instead of the single one used in the previous examples. We consider two cases,
in which the vertical offset between the seeds is $4 \, \textrm{mm}$ or
$8 \, \textrm{mm}$; their horizontal offset is $4 \, \textrm{mm}$.

\begin{figure}
  \centering
  \includegraphics[width=\textwidth]{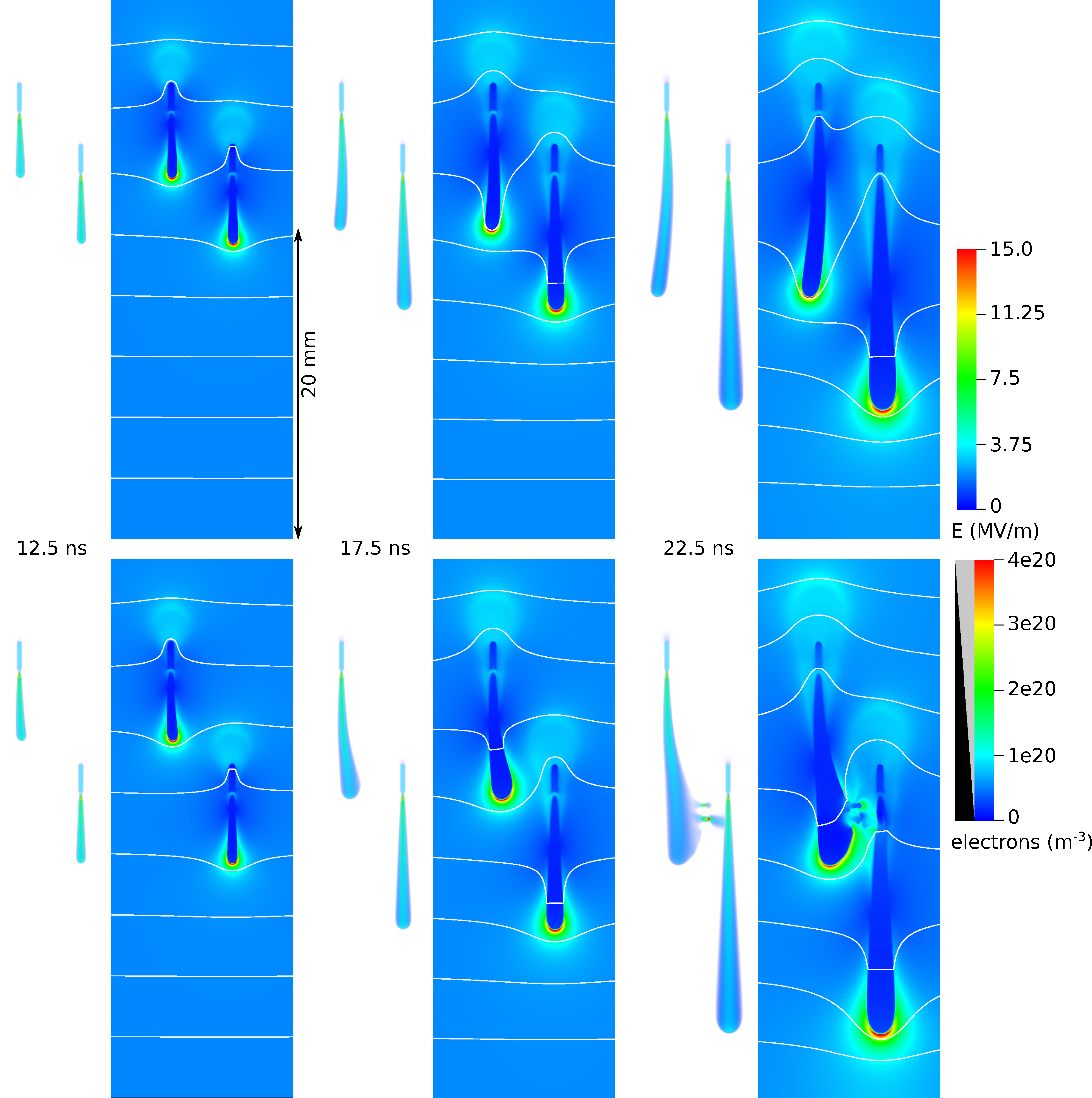}
  \caption{Evolution of two interacting positive streamers. The streamers have a
    vertical offset of $4 \, \textrm{mm}$ (top row) or $8 \, \textrm{mm}$
    (bottom row) and a horizontal offset of 4~mm. At each indicated time a 3D volume rendering of the electron
    density together with a cross section of the electric field are shown. The
    white equipotential lines are spaced by $4 \, \textrm{kV}$.}
  \label{fig:two-str-result}
\end{figure}

Figure \ref{fig:two-str-result} shows the time evolution of the electron density
and the electric field for both cases, with equipotential lines indicated at
steps of $4 \, \textrm{kV}$. With the smaller vertical offset, the streamers
repel, whereas they attract with the larger offset. This be explained by looking
at the equipotential lines. For the case with the smaller vertical offset, the
lower streamer bends equipotential lines downwards. This reduces the electric
field in which the upper streamer propagates. The reduction is smaller farther
away from the lower streamer, which causes the upper streamer to bend outwards.

For the case with the larger vertical offset, another effect becomes important.
Both streamers are \emph{in total} electrically neutral (as well as the seeds
they originate from). Their bottom/positive end therefore bends equipotential
lines downwards, whereas their upper/negative end bends them upwards. With
sufficient vertical offset between the streamers, the equipotential lines
between them are therefore compressed. This means there is an increased electric
field between them, so that they attract. In summary, positive charged streamer
heads repel, whereas a positive streamer head is attracted to a negatively
charged streamer tail. Finally, notice how in both cases the bottom streamer
propagates almost straight down, whereas path of the upper streamer is bent.

\section{Conclusions and outlook}
\label{sec:conclusions}

We have presented Afivo-streamer, an open-source plasma fluid model for 2D,
cylindrical and 3D simulations of streamer discharges. The model makes use of
the Afivo framework~\cite{Teunissen_afivo_arxiv} to provide adaptive mesh
refinement, a geometric multigrid Poisson solver and OpenMP parallelization. For
robustness, the fluid model is of the drift-diffusion-reaction type in
combination with the local field approximation. We have described the numerical
implementation of Afivo-streamer, discussing also the refinement criterion. The
model's capabilities have been demonstrated with 3D examples of long streamers
in undervolted gaps in pre-ionized nitrogen at
$1 \, \textrm{bar}$. The first example showed how stochastic background
ionization affects streamer propagation and branching. The second demonstrated
how a streamer interacts with preionized patches, in which it slows down and
loses much of its field enhancement. The third example showed how streamers can
attract or repel each other, depending on their relative position. These
simulations used up to $10^8$ grid cells, and all ran within a day.
A uniform grid with the same resolution would have required $4\cdot 10^{12}$ grid cells.

Future work will focus on the effects of photoionization, which was not included
here, but is important for discharges in air. A numerical challenge is the
inclusion of curved electrodes and dielectrics. This is planned for a future
version of Afivo-streamer, and requires not only modification of the underlying
Poisson solver as in~\cite{Celestin_2009}, but also of the fluid model around
the curved boundaries.

\ack{} JT acknowledges support by postdoctoral fellowship 12Q6117N from Research
Foundation -- Flanders (FWO).

\section*{References}
\bibliographystyle{unsrt}
\bibliography{afivo_streamer}

\end{document}